\documentstyle[pre,aps,twocolumn]{revtex}
\begin{document}
\draft
\title{Response to Comments on ``Simple Measure for Complexity''}

\author{J.S.\ Shiner\thanks{Author to whom correspondence should be addressed. 
Bergacher 3, CH-3325 Hettiswil, Switzerland. 
Phone: +41 / 34 / 411 02 43.  Electronic address:  
shiner@alumni.duke.edu}}
\address{Abteilung Nephrologie, Inselspital, CH-3010 Bern, Switzerland}
\author{Matt Davison\thanks{Electronic address:  mdavison@julian.uwo.ca}}
\address{Department of Applied Mathematics, University of Western 
Ontario, Western Science Centre, London, Ontario, Canada N6A 5B7}
\author{P.T.\ Landsberg\thanks{Electronic address: ptl@maths.soton.ac.uk}}
\address{Faculty of Mathematical Studies, University of Southampton, 
Southampton SO17 1BJ, UK}
\date{\today}
\maketitle

\begin{abstract}
We respond to the comment by Crutchfield, Feldman and Shalizi and that
by Binder and Perry, pointing out that there may be many maximum
entropies, and therefore ``disorders'' and ``simple complexities''.
Which ones are appropriate depend on the questions being addressed.
``Disorder'' is not restricted to be the ratio
of a nonequilibrium entropy to the corresponding equilibrium entropy;
therefore,  ``simple complexity'' need not vanish for all equiibrium
systems, nor must it be nonvanishing for a nonequilibrium system.
\end{abstract}

\pacs{05.20.-y, 05.90.+m}

We are pleased that our contribution on a ``simple measure for complexity''
\cite{sdl} (hereafter referred to as SDL) is of sufficient interest to
have generated two comments, one by Crutchfield, Feldman and Shalizi
\cite{CFScomment} (CFS) and another by Binder
and Perry \cite {BPcomment} (BP).  In SDL we
proposed
\begin{equation}
{\Gamma}_{\alpha \beta} \equiv {\Delta}^{\alpha} {\Omega}^{\beta},
	\label{eq1}
\end{equation}
\begin{equation}
	\Delta \equiv S/S_{max},\ \Omega \equiv 1 - \Delta.
	\label{eq2}
\end{equation}
as a ``simple measure for complexity''.  $\alpha$ and $\beta$ are
(constant) parameters, $S$ is the Boltzmann-Gibbs-Shannon entropy
\cite{Shannon1948}, and $S_{max}$, the maximum entropy.  $\Delta$ was
introduced earlier by one of us as a measure for disorder, and
$\Omega$ is referred to as ``order'' \cite{Landsberg,commas}.

CFS raise several points:
\begin{itemize}
	\item[I.] Since $S_{max}$ is the equilibrium entropy, $\Delta$ and
${\Gamma}_{\alpha \beta}$ vanish for all equilibrium systems, and
neither can ``distinguish between two-dimensional Ising systems at low
temperature, high temperature, or the critical temperature \ldots
[nor] \ldots between the many different kinds of organization observed
in equilibrium.''
	\item[II.] ${\Gamma}_{\alpha \beta}$ ``is over-universal in the sense
that it has the same dependence on disorder for structurally distinct
systems.''
	\item[III.] The ``statistical complexity'' $C_{\mu}$ of
one-dimensional
spin systems \cite{CrutchfieldandFeldman1997} is not the same as the
entropy of noninteracting spins.  The identification in
SDL of $C_{\mu}$ with the disorder of noninteracting spins is
incorrect.
\item[VI.] SDL mentions ``thermodynamic depth''
\cite{LloydandPagels1988} as a complexity measure with a convex
dependence on disorder, whereas Crutchfield and Shalizi
\cite{CrutchfieldandShalizi1998} have shown that it is an increasing
function of disorder.
\end{itemize}
The comment of BP is more specific:
${\Gamma}_{\alpha \beta}$ does not capture all aspects of complexity;
in particular,  ``it does not describe the transition from regular
to indexed languages observed at the period-doubling accumulation
points of quadratic maps.''

We welcome the two comments, as well as the opportunity to respond to
them and clarify the work presented in SDL. Let us first note that we
have only a limited interest in terminology, and if someone does not
like our use of the word ``complexity'' for the expression defined in
eq.~\ref{eq1}, let them call it the ``lambda-function'' or invent
another term.  The important thing is to have a clear definition of
the terms one is using.  For this reason, we will mostly refrain from
calling ${\Gamma}_{\alpha \beta}$ ``complexity'' in this response.

\begin{itemize}

\item[{CFS I.}] CFS have understood $S_{max}$ to {\em always} be
the equilibrium entropy, from which it follows that $\Delta$ and
${\Gamma}_{\alpha \beta}$ vanish for all equilibrium systems.  This is
a misinterpretation of SDL, perhaps due to our
choice of a nonequilibrium system to illustrate the case where the
entropy of the equiprobable distribution may not be the appropriate
$S_{max}$ and our statement that one {\em can} interpret
${\Gamma}_{\alpha \beta}$ as the product of ``order'' and ``distance
from equilibrium''.  We did not write that ``$S_{max}$ is taken to be
the equilibrium entropy of the system \ldots  for {\em all} \ldots
systems.''  \cite{CFScomment}.  Neither ``disorder'' $\Delta$ nor
${\Gamma}_{\alpha \beta}$ is restricted to this interpretation.  A
perusal of our other work \cite{Landsberg,Shiner1996,SDhier,SDLpsych}
will yield examples additional to those in SDL where $S_{max}$ is not
the equilibrium entropy of a nonequilibrium system.

In fact, it is possible to have more than one $S_{max}$, depending on
the question(s) being addressed.
\begin{itemize}
\item[(a)] Take the entropy of the universe as largely due to the
black body radiation background.  The maximum conceivable entropy can
be constructed by taking all the matter in the universe to make one
black hole, yielding a very small ``disorder'' (see, {\em e.g.}
\cite{PTLbb}).  In a sense that is an ultimate equilibrium entropy.
\item[(b)] The absolute maximum entropy possible is usually taken to
be the that of the equiprobable distribution.
\item[(c)] For a nonequilibrium system, one could take the entropy of
the equilibrium system with the same number of particles, total
energy, \ldots as the maximum entropy \cite{sdl,ideal}.
\item[(d)] The one-dimensional Ising system (two-state spins, only nearest
neighbor interactions) provides a simple example where
different $S_{max}$'s are appropriate for answering different
questions.  The entropy is a function of
the interaction parameter $J$, the external field $B$ and temperature
$T$: $S(B,J,T)$.  The case of vanishing external
field and vanishing interaction yields the equiprobable distribution
and the absolute maximum entropy: $S(B=0,J=0,T)$ \cite{hightemp}.  The
absolute ``disorder'', that with reference to $S(B=0,J=0,T)$, is then
\begin{equation}
{\Delta} = S(B,J,T)/S(B=0,J=0,T).
	\label{eq3}
\end{equation}

How much of the reduction of $S(B,J,T)$ compared to $S(B=0,J=0,T)$ is
due to the interaction between spins?  To answer this question we find
the maximum of $S(B,J,T)$ with respect to $J$ under the condition of
constant net magnetization $M$ (since, even for the paramagnet, the
entropy varies with $M$).  As expected, the entropy is maximum in the
case of vanishing interaction, $J=0$.  Thus, the maximum entropy under
the constraint of constant net magnetization is
$S({B}_{0},J=0,T)$, where ${B}_{0}$ is the value of the external field
such that $M({B}_{0},J=0,T)=M(B,J,T)$.  We now have a second
``disorder'':
\begin{equation}
{\Delta}_{J=0}=S({B}_{0},J=0,T)/S(B=0,J=0,T).
	\label{eq4}
\end{equation}
This is the absolute ``disorder'' of the paramagnet with the same net
magnetization as the Ising system with nonvanishing $J$.  Since we have
three entropies, $S(B,J,T) \le S({B}_{0},J=0,T) \le S(B=0,J=0,T)$, we
can introduce a third ``disorder'':
\begin{equation}
\widehat{\Delta} = S(B,J,T)/S({B}_{0},J=0,T),
	\label{eq5}
\end{equation}
which is the ``disorder'' of the Ising system with respect to the
paramagnet with the same net magnetization.  The three ``disorders''
are related by ${\Delta} = \widehat{\Delta}{\Delta}_{J=0}$.
\end{itemize}

The point is that there are many possible $S_{max}$'s, and therefore
``disorders'', ``orders'' and ``complexities'' ${\Gamma}_{\alpha
\beta}$, even for equilibrium systems.  Which one(s) are appropriate
depends on the question(s) being addressed.  It is not in general true
that $\Delta$ is identically $1$ for equilibrium systems; therefore
neither ``order'' nor ``complexity'' must vanish at equilibrium.  When
CFS write that as a consequence of $S_{max}$'s being taken as the
equilibrium entropy (for nonequilibrium systems) neither $\Delta$ nor
${\Gamma}_{\alpha \beta}$ can ``distinguish between two-dimensional
Ising systems at low temperature, high temperature, or the critical
temperature \ldots  [n]or \ldots  between the many different {\em kinds} of
organization observed in equilibrium'', this is an overly restrictive
interpretation of $\Delta$ and ${\Gamma}_{\alpha \beta}$.

For a paramagnet ``pumped out of equilibrium'', one could interpret
$S$ and $S_{max}$ as the nonequilibrium entropy and the equilibrium
entropy \emph{under the appropriate constraints} \cite{ideal},
respectively.  However, here again, CFS misinterpret our work.  They
wish to argue that since the pumped state is out of equilibrium, we
would assign a nonzero level of complexity to this state.  This is not
true for this simple case of a paramagnet, for which the entropy can
be written simply in terms of the total number of spins and the net
magnetization.  The key is to realize that the appropriate constraints
here are just the total number of spins and the net magnetization;
otherwise the nonequilibrium entropy could be greater than the
equilibrium entropy.  Since the total number of spins and the net
magnetization must then be the same for the equilibrium and the
nonequilibrium case, the entropies are the same, and we have maximum
``disorder'' and vanishing ${\Gamma}_{\alpha \beta}$.  Incidentally,
we have never maintained that $1 - \Delta$, where $\Delta$ is the
ratio of a nonequilibrium entropy to the corresponding equilibrium
entropy, could distinguish different equilibrium distributions.  To do
this, one needs some of the various equilibrium disorders, as pointed
out above.

\item[{CFS II.}] CFS write that ${\Gamma}_{\alpha \beta}$ is
``over-universal in the sense that it has the same dependence on
disorder for structurally distinct systems.''  We assume they mean
that ${\Gamma}_{\alpha \beta}$ always has the same dependence on
``disorder'' (given $\alpha$ and $\beta$).  We agree with this as far
as it goes; it follows from the definition of ${\Gamma}_{\alpha
\beta}$.  It is rather superficial though, and the question arises as
to which ${\Gamma}_{\alpha \beta}$ and which ``disorder'' are meant.
If ${\Gamma}_{\alpha \beta}$ is calculated from one ``disorder'' and
its dependence on another ``disorder'' investigated, ${\Gamma}_{\alpha
\beta}$ may well have a variable dependence on ``disorder.''  In Fig.~4
of SDL, we investigated ${\Gamma}_{11, J=0}$ as a function of
$\Delta$ for one-dimensional Ising systems.  This relation varies with
$J$.  Thus, it is not generally true that ${\Gamma}_{\alpha \beta}$
``has the same dependence on disorder for structurally distinct
systems.''  One has to be careful to clearly state which ``disorder''
and which ``complexity'' one is dealing with.  If one does so, then
${\Gamma}_{\alpha \beta}$ may show different dependencies on
``disorder'' for ``structurally distinct systems'' and is not
``over-universal'' in the sense used here.

Our calculation of ${\Gamma}_{11, J=0}$ as a function of $\Delta$
is analogous to Crutchfield and Feldman's
\cite{CrutchfieldandFeldman1997} calculation of ``statistical
complexity'' $C_{\mu}$ and ``excess entropy'' $E$, or ``effective
measure complexity'' \cite{Grassberger}, again for one-dimensional
Ising systems.  In our interpretation of their results, they found
$C_{\mu}$, to within a multiplicative constant, to be ${\Delta}_{J=0}$
(we will return to this point below), and $E$, again to within a
multiplicative constant, to be ${\Delta}_{J=0} - \Delta$.  They then
investigated the dependence of ${C}_{\mu}$ and $E$ on $\Delta$ and
found that these dependencies vary with $J$.  From an
``order''-``disorder'' point of view, what they have investigated is
the dependence of ``disorder'' under one set of conditions ($J = 0$)
on ``disorder'' under another set of conditions ($J \neq 0$), or in
the case of $E$, the difference between these two ``disorders'' on one
of the ``disorders''.

\item[{CFS III.}] In SDL we identified $C_{\mu}$ for the
one-dimensional Ising system, to within a multiplicative constant,
with the ``disorder'' of that system in the absence of interactions
between spins.  CFS object to this on two grounds, the first of which
is dimensional inconsistency.  This is not the place to get into a
discussion of the proper units for entropy; let us just reiterate
-- to within a multiplicative constant.  More seriously, CFS maintain
that we ``conflate'' the definition of $C_{\mu}$ with eq.~(8) of
\cite{CrutchfieldandFeldman1997}, which is only valid within a limited
range.  Actually, we were not referring to that equation to identify
$C_{\mu}$, but rather to identify the excess entropy $E$.  Be that as
it may, our identification of $C_{\mu}$ with ${\Delta}_{J=0}$
in SDL is restricted to one-dimensional spin systems, which is what
they treat in \cite{CrutchfieldandFeldman1997} and we treat in SDL. We
were not ``conflating'', but referring to their results for
one-dimensional spin systems.  They disagree with this, too, when they
say that although $C_{\mu} = H(1)$, $H(1)$ is not the same as the
entropy of noninteracting spins.  However, on pg.  1240 of
\cite{CrutchfieldandFeldman1997} they write: ``For a NN system, Eq.
(6) is equivalent to saying that $C_{\mu} = H(1)$, the entropy
associated with the value of one spin.''  Earlier, pg.  1239, they
used the phrase ``isolated-spin uncertainty $H(1)$''.  Since an
isolated spin can not be subject to interactions with neighboring
spins, to our reading, they themselves have stated that $H(1)$ is the
entropy of a spin subject to no interactions, and thus, to within a
multiplicative constant, just ${\Delta}_{J=0}$.  (Note that according
to \cite{CrutchfieldandFeldman1997} the identification of $C_{\mu}$
with $H(1)$ breaks down for the paramagnetic case and the high
temperature limit; we exclude these cases, too, of course.)

\item[{CFS IV.}] Does ``thermodynamic depth''
\cite{LloydandPagels1988} show a convex dependence on ``disorder'', as
we stated in SDL, or does it increase monotonically with ``disorder''
\cite{CrutchfieldandShalizi1998}?  Our statement was based on the
original exposition by Lloyd and Pagels \cite{LloydandPagels1988} and
other discussions (e.g.  \cite{Wackerbaueretal1994}).  The results of
Crutchfield and Shalizi \cite{CrutchfieldandShalizi1998} would indeed
seem to indicate that thermodynamic depth is a monotonically
increasing function of ``disorder'', given their insistence that the
choice of states to be made should be the ``causal states'' of
``$\epsilon$ - machines''.  In particular, they object to 
``judiciously redefining'' (pg. 277) the ``appropriate set'' of 
macroscopic states.  However, we believe the
situation may not be so simple.  First of all, they write
\cite{CrutchfieldandShalizi1998} (pg. 278): ``It is certainly not desirable to
conflate a process's complexity with the complexity of whatever
apparatus connects the process to the variables we happen to have
seized upon as handles''.  This argument ignores the fact that the
only access we have to real systems is through measurement.  The
situation would seem to be reminiscent of the endo-exophysics
distinction (see, e.g., \cite{Endoexo}), at least superficially.
Crutchfield and Shalizi take more of a endophysical point of view,
while it would seem that Lloyd and Pagels take a more exophysical
approach.  Along a similar vein, the argument of Crutchfield and
Shalizi seems to ignore the problem
of frames of reference.  For example, Andresen and Gordon
\cite{Andresengordon} and Spirkl and Ries \cite {Spirklries} have
shown that a necessary condition for minimum entropy production in a
continuous time system is a constant rate of entropy production in
{\em eigen} time.  This can be described as the instantaneous internal
time scale of the system and is different from clock or wall time
except for linear systems.  In other words, for nonlinear systems the
rate of entropy production will not be constant for an external
observer, but only to the (nonlinear) system as it sees itself.  In
any case, ``thermodynamic depth'' would seem to be a convex measure of
``complexity'' in some cases.  ``Back of the envelope'' calculations
of ``thermodynamic depth'', taken as the difference between a
coarse-grained entropy and a fine-grained entropy \cite{abuse}, for a
one-dimensional Ising ferromagnet indicate a convex dependence on
``disorder''.  However, this is not the case for an antiferromagnet
with sufficiently negative $J$.  Thus, ``thermodynamic depth'' may not
qualify as either a convex or a monotonic complexity measure, or it
may be either depending on the particulars of the system being
investigated.

\item[{BP.}] We agree with Binder and Perry (BP) that results obtained
on the basis of ${\Gamma}_{\alpha \beta}$ should be carefully
interpreted and complemented with results based on other measures, if
possible.  However, we are not so sure that all of their statements
are completely accurate.  For example, they argue that ``effective
measure complexity'' will ``[c]ertainly \ldots  pick up the
non-regularity of a language'', but that our measure will not in the
logistic map.  However, a comparison of Fig.\ 10 of
\cite{Wackerbaueretal1994} and Fig.\ 3 of SDL show that, as stated in
SDL, ``major maxima as well as less major ones occur at the same
values of $r$'', although the relative values of the maxima differ.

Perhaps the main point of contention is that BP desire a``complexity''
measure which can become infinite, whereas we purposefully constructed
the measure in SDL so that it would not have this property (for
$\alpha, \beta \ge 0$).  Our reasoning is similar to that which argues
that $S/{S}_{max}$ is, for certain purposes, a ``better'' measure
for``disorder'' than is the entropy $S$ \cite{Landsberg}.

It would also seem that BP, like CFS, may have taken the definition of
${\Gamma}_{\alpha \beta}$ too literally in that they may not have
realized that there may be several different ${S}_{max}$'s, and
therefore $\Delta$'s, and therefore ${\Gamma}_{\alpha \beta}$'s for a
given system.

Nonetheless, we would like to reiterate that we stated only that
${\Gamma}_{11}$ behaves {\em similarly} to ``effective measure
complexity'' for the logistic map, that it was not clear to us why
this is so, and we do not know the breadth of systems for which this
will be the case.
\end{itemize}

There is a plethora of proposed complexity measures
in the literature, all trying to capture some aspects of what we mean
when we say that something is complex.  Among ones that we ourselves
find attractive are ``thermodynamic depth'', ``statistical complexity''
and ``effective measure complexity''.  However, none of these nor any
of the others capture all aspects of ``complexity''.  This is made
explicit by the statement by CFS ``that a useful role for statistical
complexity measures is to capture the structure -- patterns,
organization, regularities, symmetries -- intrinsic to a process''.
We have nothing against this statement, unless one interprets ``a
useful role'' as `` the only useful role'', or one means that a measure
of complexity must be a statistical complexity measure.  There are
many useful roles for complexity measures. \cite{wf} Perhaps at some time a
consensus will arise; but until that time, we believe that there is a
need for various approaches to complexity.

The situation becomes even more confused, when one realizes that even
seemingly ``exact'' measures such as ``statistical complexity'' and
``effective measure complexity'' are not uniquely defined: ``For
higher dimensional systems, e.g., spins in 2D, there are several ways
to define $E$ and $C_{\mu}$.''  \cite{CrutchfieldandFeldman1997} (pg. R1242)
 Thus,
we believe there is a place for simple measures of complexity.  The great
advantage of ${\Gamma}_{\alpha \beta}$, as noted in SDL, is that it is
available for systems where much less information is available than is
necessary to calculate some other measures, such as $E$ and $C_{\mu}$.

We do not claim any ``universality'' for ${\Gamma}_{\alpha
\beta}$ though, and think that one
should examine several possible complexity measures to
get a handle on the various things which can be meant by saying a
system or a process is complex.

This work was supported in part by grant no. 31-42069.94 from the 
Swiss National Science Foundation.


\begin{thebibliography}{990}
\bibitem{commas}When referring to specific measures of disorder, order
  or complexity, we enclose these words in quotation marks to
  distinguish the measures from the general concepts.
\bibitem{hightemp}The maximally random state is also reached in the
	high temperature limit.
\bibitem{abuse} In the ``back of the envelope'' calculation we may abuse
Lloyd and
Pagels's concept of ``thermodynamic depth''.  They are concerned with
the entropies of processes, whereas we worked with state entropies.
\end{thebibliography}
 \end{document}